\documentclass[prl,aps,twocolumn,showpacs,floatfix]{revtex4}
\usepackage{epsfig}
\usepackage{graphicx}
\usepackage[dvips]{color}

\newcommand{\beq}{\begin{displaymath}}
\newcommand{\eeq}{\end{displaymath}}
\newcommand{\ben}{\begin{equation}}
\newcommand{\een}{\end{equation}}
\newcommand{\bea}{\begin{eqnarray}}
\newcommand{\eea}{\end{eqnarray}}

\begin{document}
\author{S. Boccaletti$^{1}$,
           M. Chavez$^{2}$, A. Amann$^{3}$ and
           D.-U. Hwang$^{1}$}
\affiliation{$1$ CNR-Istituto dei Sistemi Complessi, Largo E.
Fermi, 6, Florence, Italy \\
$2$ Laboratoire de Neurosciences Cognitives et
Imagerie  C\'{e}r\'{e}brale (LENA) CNRS UPR-640 \\
H\^{o}pital de la   Salp\^{e}tri\`{e}re. 47~Bd. de l'H\^{o}pital,
75651 Paris CEDEX 13,
  France \\
  $3$ Tyndall National Institute, Cork,
Ireland}
\title{Synchronization in Complex Networks: a Reply on a recent Comment}

\begin{abstract}

We clarify a number of points raised in [Matias,
arXiv:cond-mat/0507471v2 (2005)].
\end{abstract}


\maketitle

In a previous comment on cond-mat \cite{matias}, two of our recent
publications \cite{noiuno,noidue} were compared with researches
simultaneously carried out by the Potsdam group and published in
\cite{potsdam1,potsdam2}. It was argued that the conclusions of
both sets of papers are {\em closely related}. To obviate
confusion, we will in the following point out a number of facts
which will help to better understand the differences in the
results and conclusions of the involved works.

\begin{enumerate}

\item{} The root for the improvement observed in \cite{potsdam1,potsdam2}
is the normalization of the diagonal terms of the connectivity
matrix. Such a normalization is limited in
\cite{potsdam1,potsdam2} to the optimal case $\beta=1$, whereas in
our papers it is realized {\it for all values of} $\alpha$
\cite{noiuno} {\it  and} $\theta$ \cite{noidue}. Furthermore, the
optimal condition $\beta=1$ in \cite{potsdam1,potsdam2}
corresponds to  $\alpha=0$ in \cite{noiuno} and to $\theta=0$ in
\cite{noidue}. Therefore, any further synchronizability
enhancement observed, e.g., in Figs. 2,3 of \cite{noiuno} and in
Figs. 1,2 of \cite{noidue} cannot be related to any of the
mechanisms reported in \cite{potsdam1,potsdam2}. It is therefore
not appropriate to compare the enhancement of
\cite{potsdam1,potsdam2} with respect to a non weighted
configuration to the further improvements that our original
approaches provide. For instance, the very nice analytical study
of \cite{potsdam1,potsdam2} (duly quoted by us in \cite{noiuno})
is restricted to a series of explicit hypotheses that require a
sufficiently "random" configuration to be satisfied. Our studies
manifestly show (see e.g. Fig. 1 and 2c of \cite{noidue}) that
while for Erd\"os-R\'enyi graphs the maximal synchronizability is
attained in the condition of \cite{potsdam1,potsdam2} (that is for
$\theta=0$), the interplay between hierarchy and weighting is able
to further improve this value for other network's configurations
{\it with the same property of constant total strength of input
connections}. This extra enhancement {\it cannot} be extracted by
the arguments raised in \cite{potsdam1,potsdam2}.

\item{} A weighted wiring
does not necessarily imply asymmetric connections between
elements. For instance, the weighting procedure of
\cite{potsdam1,potsdam2}  (based on powers of the node degrees
$k_i^{\beta}$) yields {\it always} (i.e. for all $\beta$) a
symmetric connection between connectivity peers (as e.g. it can
frequently occur in assortative networks). At variance, in
\cite{noidue} we are able to modulate and direct the interaction
between {\it any pair} of nodes by the parameter $\theta \in
(-1,1)$ (from a unidirectional to a bidirectional symmetrical
one), accordingly to the hierarchical (age) ordering of the nodes.
The conclusion that a weighted configuration based on a
hierarchical structure can enhance synchronization is, therefore,
original.

\item{} Our studies
\cite{noiuno,noidue} report a detailed analysis of the propensity
for synchronization in parameter spaces. Namely, in \cite{noiuno}
(\cite{noidue}) , we detail in Fig. 2 (Figs. 1,2) the behavior of
$\lambda_N/\lambda_2$ ($\lambda_N^r/\lambda_2^r$) in the parameter
space [$\alpha,B$] ($\theta$), showing that the enhancement due to
a weighting procedure based on the link loads (the age ordering)
is a generic feature in parameter space. In contrast to what was
argued in \cite{matias}, we in fact carefully considered the
dependence of our results on the network size $N$ in both our
papers. In particular, we show via Gershgorin's circle theorem
that the normalization procedure yields an upper bound for
$\lambda_N$ {\it for any network size and topology}. Notice that,
in \cite{potsdam1,potsdam2}, when the weight is selected to be
$k_i^{\beta}$, $\lambda_N$ is in general size dependent for
$\beta<0$, and one has $\lim_{N \to \infty} \lambda_N = \infty$.

\item{} The Author of \cite{matias} is incorrect when stating that the
main conclusion of \cite{noidue} is that {\it the enhancement in
synchronization is evident when the coupling direction is from the
younger to the older nodes.} Our Letter explicitly claims that one
has to combine this property (relaxed also in
\cite{potsdam1,potsdam2} for $\beta>0$) with a structure of
interconnected hubs induced by the hierarchical (age) ordering.
The second part of our Letter (as well as Fig. 2c of
\cite{noidue}), is entirely dedicated to study a series of ad-hoc
modified networks with the aim of pointing out how such an
interplay works in scale-free networks. Furthermore, it is
straightforward to show that {\it for any network size and
topology} one can construct a coupling scheme based solely on age
ordering and able to give $\lambda_N/\lambda_2=1$ (the minimal as
possible value). This can be realized by constructing a
uni-directional tree structure, by means of which one randomly
selected node forces a cascade of nodes spanning the whole
networks. Using properly normalized weights for each
uni-directional connection (i.e. normalizing the connection
strength to the in-degree of the slave node), and imposing a
zero-row sum configuration induces a triangular matrix whose
diagonal elements are (0,1,1,1,1,1,1,.....,1), giving the
following set of eigenvalues $\lambda_1=0$, $\lambda_i=1$, $i=2,
\ldots N$ \cite{note}.

\item{} It is well known that a crucial difference in topological
information content exists between the node degree and the load of
the link {\it or edge betweennes}, the former retaining {\it only}
information on the local structure of the network, the latter
providing a measure of centrality of the node \cite{load} in the
global topological structure. It is also well known (e.g. in the
study of network's community structures) that nodes with high
degrees do not necessarily have links with high loads, and links
with high loads do not necessarily connect two nodes with high
degrees. An important and original result of our studies
\cite{noiuno} is therefore that (under the assumption of input
normalization) the knowledge of global topological properties of
the network can be used to improve the propensity of
synchronization over the limit obtained in
\cite{potsdam1,potsdam2}, where only local properties were used.
This improvement is not {\em closely related} to the ideas or
conclusions of \cite{potsdam1,potsdam2}.

\end{enumerate}

In summary we have shown that the results and conclusions in Refs.
\cite{potsdam1,potsdam2,noiuno,noidue} are independent and
complementary. However, though from a completely different point
of view than that suggested by \cite{matias}, a relationship
between these works indeed exists. Taken together, the approaches
of \cite{potsdam1,potsdam2,noiuno,noidue} have undoubtedly {\it
and independently} established weighted networks as a promising
framework for the further study of synchronized behavior. While
the arguments provided in the present Manuscript demonstrate that
it is erroneous trying to deduce the conclusions of one approach
from the other, we believe that much effort will be stimulated in
the forthcoming years to investigate other ad-hoc weighted
configurations, for assessing more deeply the underlying
mechanisms for the formation of collective dynamics in all those
circumstances where the weighting features can be directly
extracted from available data.

This work was supported by MIUR-FIRB project n. RBNE01CW3M-001.


\begin{thebibliography}{99}

\bibitem{matias}
M. Matias, arXiv:cond-mat/0507471v2 (2005).

\bibitem{noiuno}
M. Chavez, D.-U. Hwang, A. Amann, H.G.E. Hentschel and S.
Boccaletti, Phys. Rev. Lett. {\bf 94}, 218701 (2005).

\bibitem{noidue}
D.-U. Hwang, M. Chavez, A. Amann and S. Boccaletti, Phys. Rev.
Lett. {\bf 94}, 138701 (2005); arXiv:cond-mat/0412728.

\bibitem{potsdam1}
A.E. Motter, C.S. Zhou and J. Kurths, Europhys. Lett. {\bf 69},
334 (2005); arXiv:cond-mat/0406207.

\bibitem{potsdam2}
A.E. Motter, C.S. Zhou and J. Kurths, Phys. Rev. E {\bf 71},
016116 (2005); arXiv:cond-mat/0502309.

\bibitem{note}
The full description of this latter weighting procedure will be
reported by us in a forthcoming paper.

\bibitem{load}
L. C. Freeman, Social Networks {\bf 1} 215 (1979); M. E. J.
Newman, Phys. Rev. {\bf E64}, 016132 (2001); U. Brandes, Journ. of
Mathematical Sociology {\bf 25}, 163 (2001);
 K.\ I.\ Goh, B.\ Kahng, and D.\ Kim, Phys. Rev. Lett. {\bf 87}, 278701 (2001).



\end{thebibliography}
\end{document}